\documentclass[sigconf]{acmart}

\copyrightyear{2025}
\acmYear{2025}
\setcopyright{cc}
\setcctype{by}
\acmConference[FSE Companion '25]{33rd ACM International Conference on the
Foundations of Software Engineering}{June 23--28, 2025}{Trondheim, Norway}
\acmBooktitle{33rd ACM International Conference on the Foundations of
Software Engineering (FSE Companion '25), June 23--28, 2025, Trondheim,
Norway}\acmDOI{10.1145/3696630.3728701}
\acmISBN{979-8-4007-1276-0/2025/06}

\usepackage{chngcntr}
\usepackage{graphicx}
\usepackage{amsmath}
\usepackage{subcaption}
\usepackage{algorithmic}
\usepackage{soul}
\usepackage{url}
\usepackage{multirow, multicol, makecell}
\usepackage{balance}
\usepackage{amsmath,amsfonts}
\usepackage{algorithmic}
\usepackage{algorithm}
\usepackage{array}
\usepackage{url}
\usepackage{textcomp}
\usepackage{stfloats}
\usepackage{url}
\usepackage{verbatim}
\usepackage{balance}
\usepackage{graphicx} 
\usepackage[breakable,skins]{tcolorbox}
\usepackage{xcolor}
\usepackage{enumitem}
\usepackage{tabularx}
\usepackage{hyperref}
\usepackage{xspace}
\usepackage{mathpazo}
\usepackage{url} 
\usepackage{xspace}
\usepackage{framed}
\usepackage{enumitem} 
\usepackage[breakable,skins]{tcolorbox}

\usepackage{booktabs}

\newcommand{\Name}{LLPut}
\newcommand{\palatino}{\fontfamily{ppl}\selectfont}

\newenvironment{promptbox}[2][]{
    \begin{tcolorbox}[title={#2},
    fonttitle={\palatino\bfseries}, 
    enhanced jigsaw, 
    colbacktitle=black,
    arc=2pt,
    opacityframe=0,
    boxrule=0.4mm,
    opacityframe=1,
    colback=green!10!white,
    breakable]
}{\end{tcolorbox}}

\title{LLPut: Investigating Large Language Models for Bug Report-Based Input Generation}

\begin{document}

\balance

\author{Alif Al Hasan}
\affiliation{%
  \institution{Jahangirnagar University}
  \city{Dhaka}
  \country{Bangladesh}
}
\email{alif.stu2017@juniv.edu}

\author{Subarna Saha}
\affiliation{%
  \institution{Jahangirnagar University}
  \city{Dhaka}
  \country{Bangladesh}
}
\email{subarna.stu2019@juniv.edu}

\author{Mia Mohammad Imran}
\affiliation{%
  \institution{Missouri University of Science and Technology}
  \city{Rolla, Missouri}
  \country{USA}
}
\email{imranm@mst.edu}

\author{Tarannum Shaila Zaman}
\affiliation{%
  \institution{University of Maryland Baltimore County}
  \city{Baltimore, Maryland}
  \country{USA}
}
\email{zamant@umbc.edu}

\begin{abstract}
Failure-inducing inputs play a crucial role in diagnosing and analyzing software bugs. Bug reports typically contain these inputs, which developers extract to facilitate debugging. Since bug reports are written in natural language, prior research has leveraged various Natural Language Processing (NLP) techniques for automated input extraction. With the advent of Large Language Models (LLMs), an important research question arises: how effectively can generative LLMs extract failure-inducing inputs from bug reports? In this paper, we propose \Name{}, a technique to empirically evaluate the performance of three open-source generative LLMs—LLaMA, Qwen, and Qwen-Coder—in extracting relevant inputs from bug reports. We conduct an experimental evaluation on a dataset of 206 bug reports to assess the accuracy and effectiveness of these models. Our findings provide insights into the capabilities and limitations of generative LLMs in automated bug diagnosis.
\end{abstract}



\maketitle

\begin{sloppypar}

\section{Introduction}
Bug reports serve as the primary medium through which users and testers communicate software issues to developers \cite{10.1145/2652524.2652541}. These reports typically include a detailed natural language description of the problem, aiding developers in reproducing and diagnosing the bug. As a result, bug reports are a crucial component of software maintenance. Their content is not only essential for developers in maintaining software systems but also forms the foundation for automated tools that assist in the complex tasks of bug detection and resolution \cite{10.1145/2652524.2652541}.
Whenever developers receive a bug report, the first step is to reproduce the bug \cite{Adreas:Zeller, zaman2025genderdynamicssoftwareengineering}. It is necessary to reproduce the bug to be confirmed and to observe the behavior of the bug. Bug reproduction also enables further analysis of the issue \cite{Casey:1992:CPD:144167}.

To reproduce a bug, developers need the failure-inducing input \cite{10.1145/3106237.3106266, 10.1145/3476883.3520207}. These inputs are test cases or commands that cause a software failure during execution \cite{10.1145/347636.348938}. Developers also rely on inputs to diagnose and localize bugs. Ideally, bug reports clearly specify failure-inducing input commands. However, unstructured and ambiguous information in bug reports often makes them difficult to use \cite{8811942}. Additionally, many bug reports describe inputs in natural language rather than technical terms. For example, instead of writing the command \textit{chmod filename} to change a file’s mode, a reporter might say, \textit{change the file’s permission mode.} This use of natural language makes it harder for new developers to extract the necessary inputs from bug reports.

Since bug reports are written in natural language, researchers have conducted numerous studies \cite{lota5060080recent} using NLP \cite{du2024llmsassistnlpresearchers} techniques to analyze bug reports and extract relevant information. This is the first study to systematically evaluate the effectiveness of LLMs in extracting input commands from bug reports. Recently, LLMs have assisted NLP researchers in various fields by analyzing human-written documents and extracting information from them \cite{du2024llmsassistnlpresearchers}.

The task of extracting \textit{input commands} from bug report can be categorized as information extraction task. In recent years, the use of generative LLMs for information extraction has been popular~\cite{zhang2025survey, sainz2023gollie, alaofi2024generative, deng2024information, xu2024large}, including software engineering tasks~\cite{mo2024c, li2023codeie}. They have been used in a zero-shot, one-shot, and few-shot manner to extract specific tasks. In software engineering, while LLMs have been used for tasks such as code summarization, bug localization, code generation, and vulnerability detection~\cite{ahmed2022few, ahmed2024automatic, hossain2024deep, yang2024large, lu2024grace, akuthota2023vulnerability, mo2024c, li2023codeie, gu2023llm}, extracting input commands in bug reports has been not previously performed. This gap inspired us to investigate the effectiveness of the LLMs in extracting manual commands from bug reports. Therefore, we propose a tehcnique \Name{} to analyze LLM's performance in extracting inputs from bug report.

\begin{figure}[htb]
\centering
\includegraphics[scale=0.55]{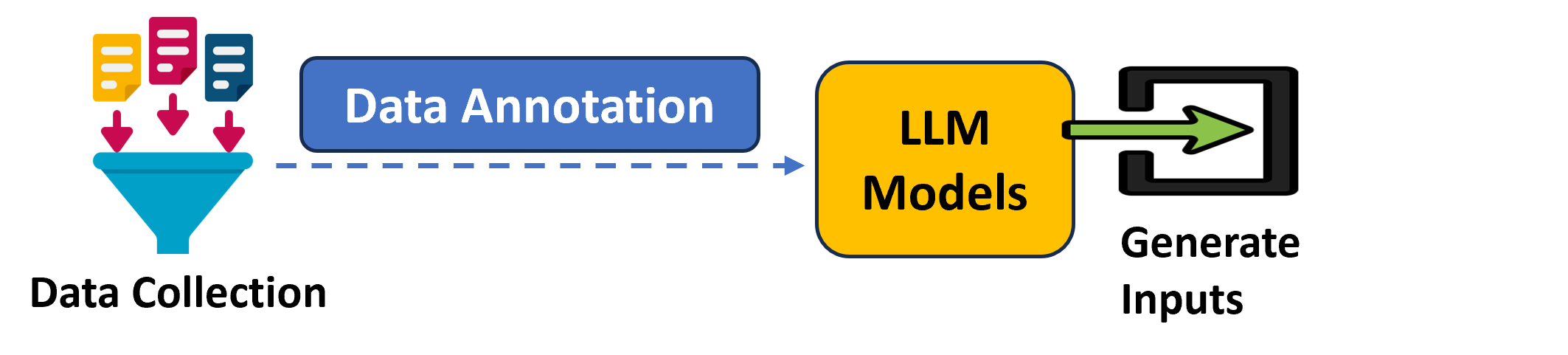}
\caption{Overview of \Name{} Methodology}
\label{fig:overview}
\end{figure}

Our proposed methodology, \Name{}, consists of two main components: (i) dataset preparation and (ii) application of various LLM models for input extraction. Figure \ref{fig:overview} provides an overview of \Name{}'s methodology. We begin by collecting and annotating bug reports to construct a structured dataset. As an initial baseline, we apply a BERT-based NLP technique to evaluate its effectiveness in extracting failure-inducing inputs. Subsequently, we employ three generative LLM models—LLaMA \cite{touvron2023llama}, Qwen \cite{bai2023qwen}, and Qwen-Coder \cite{hui2024qwen2}—to evaluate the performance of LLMs in extracting inputs from bug reports.
Our dataset, codes,  annotation instructions and prompts are available in this repo \cite{dataset}.



\section{Dataset}
To analyze the performance of the models in generating inputs from bug reports, we first collect and annotate a dataset of bug reports. 
\subsection{Dataset Collection}  
In this work, we focus on the Linux coreutils project, which includes 68 distinct utility applications \cite{coreutils} and offers a diverse range of functionalities. We collect bug reports from the online open source bug repository, \textit{Red Hat Bugzilla - Bug List} \cite{redhat}. We filter bug reports by selecting \textit{Fedora} and \textit{Red Hat Linux} as the product categories and \textit{coreutils} as the target component. This filtering yields a total of 779 bug reports. We then retrieve these reports using the \textbf{REST API} and parse the data into two separate CSV files: one containing bug descriptions with their corresponding IDs and another containing IDs along with all other relevant attributes.
In the future, we plan to expand our dataset by collecting data from various web-based applications and from other different open-source bug repositories.

To refine the dataset, we manually review the bug reports and exclude those that merely announce new \textit{coreutils} releases. After this filtering process, we obtain a final dataset of 753 bug reports. Table \ref{tab:sample-data} presents a sample from our dataset.
\begin{table*}
    \centering
    \caption{Sample Data}
    \label{tab:sample-data}
    \renewcommand{\arraystretch}{1.1}  
    \begin{tabular}{|p{1cm}| p{1.5cm}| p{8cm} | p{5cm} |}  
        \hline
        \textbf{id} & \textbf{title} & \textbf{Description} & \textbf{Input Command(s)} \\ 
        \hline
        2296026 & 
        sort \verb|-V| order differs from \verb|sort| when there are no numerics & 
        When sorting two lines with no numeric parts, \verb|sort -V| differs from \verb|sort|. \newline
        [...]
        \textbf{How reproducible:} Always \newline
        Note: The results are the same (as expected) if lines ``a'' and ``abc'' are sorted 
        (i.e., without the ``-xyz''): \newline
        \verb|printf "a\nabc\n" | sort -V| \newline
        a \newline
        abc [...] &
        1. \verb|printf "a-xyz\nabc-xyz\n" | sort| \newline
        2. \verb|printf "a-xyz\nabc-xyz\n" | sort -V| \\
        
        \hline
        2248503 & 
        uname \verb|-i| doesn't work & 
        \verb|coreutils-9.1-12.fc38.x86_64| \newline
        \verb|[root@buildserver:~]$ uname -i| \newline
        \verb|unknown| \newline
        This is breaking scripts expecting to get \verb|x86_64|. &
        1. \verb|uname -i| \\
        
        \hline
        749704 & 
        date cannot easily produce lower case formatted output & 
        Suppose one wants lower case abbreviated day-of-week names. By default, \verb|%a| produces capitalized names, e.g., Fri. Using \verb|%^a| correctly produces upper case. Using \verb|%#a| unfortunately produces the same as \verb|%^A|, i.e., FRI. Quite clearly, FRI is not the opposite case of Fri. Instead of an 'opposite case' modifier 
        [...] 
        &
        1. \verb|date +%a| \newline
        2. \verb|date +^a| \newline
        3. \verb|date +#a| \newline
        4. \verb|date +%/a| \\
        
        \hline
    \end{tabular}
\end{table*}

\subsection{Dataset Annotation}  
Two authors manually annotate a randomly selected 250 reports from the refined dataset of 753 reports and discuss until they reach to consensus. They further discard reports with no or vague descriptions (i.e., bug reports had nothing beyond an imprecise title, or any details that could help a user understand the bug). Afterward, we end up with a subset of 206 reports for further analysis. We give annotators the following instructions:

\begin{quote}
{\em Read the entire bug report thoroughly before beginning annotation. Pay attention to both explicit statements and implicit cues that might indicate the exact conditions which triggered the anomaly. For each bug report, determine the specific commands/instructions and in which order triggered that state.
}
\end{quote}

The detailed instruction is available in the project repository. The annotators carefully examine each report to identify the exact sequence of commands needed to reproduce the reported bugs.

Annotators note that many bug reports lack proper structure or omit essential details for understanding the issue. In these cases, annotators thoroughly read each report and extract the necessary information, including the expected outcome, observed outcome, and steps to reproduce the bug. Some reports only describe the encountered issue and how the user reached that state. The annotators analyze these cases, interpret the context, and reconstruct the steps to ensure accurate replication.

After annotation, We find that in 149 bug reports, input commands or test cases exist. 54 cases did not contain input commands or test cases. We mark them as `None'. And in 3 cases annotators were uncertain as the actual commands were not explicit. We also mark them as `None'.

\section{Experimental Evaluation} In this research work, we evaluate \Name{}, by considering two research questions: \\
\textbf{RQ 1:} How effective is BERT in accurately extracting user input commands from unstructured bug reports?\\
\textbf{RQ 2:} How effective is generative LLMS in extracting inputs from bug report?\\

\subsection{Evaluation Metric}

As \textit{input commands} are needed to exactly match in order to run correctly, we compare the NLP (BERT model) extracted input commands and  Genrative LLM extracted input commands against human-annotated ones using BLEU score (Bilingual Evaluation Understudy)~\cite{papineni2002bleu}. BLEU score is a standard metric in machine translation to evaluate text sequence similarity. In the software engineering domain, BLEU score is often used to evaluate the quality of generated bug reports, code comments, and code summarization.

\subsubsection{BLEU-score}: 
The BLEU score is a standard metric used to assess the quality of machine-generated text by comparing it to human-written reference text~\cite{papineni2002bleu}. It measures the similarity between the generated and reference texts by evaluating the overlap of n-grams. A higher BLEU score indicates a closer resemblance to the reference, making it particularly useful for tasks like machine translation and text generation. The BLEU score is computed using the following formula:  

$$
BLEU = BP \cdot \exp\left(\sum_{n=1}^{N}w_n \cdot \log(p_n)\right)
$$
where, $N$ represents the maximum n-gram order considered in the calculation;
$p_n$ is the \textit{precision score} for n-grams of size $n$, reflecting the proportion of matching n-grams between the generated and reference texts;
$w_n$ is the weight assigned to each n-gram level, typically set to $\frac{1}{N}$ to ensure equal importance across all n-gram orders; and
$BP$ denotes the \textit{brevity penalty}, applied to discourage excessively short outputs. It equals 1 if the generated text is longer than the reference and less than 1 otherwise.

\subsubsection{BLEU Score Interpretation}:
BLEU score above 0.30 generally indicates that the generated text is understandable, while score exceeding 0.50 suggests high fluency and quality~\cite{denkowski2010choosing}. This scale has been widely applied in various fields~\cite{google_translate_automl_evaluate, lavie-2011-evaluating, Seljan2012BLEUEO}, including software engineering~\cite{gao2020generating, imran2024uncovering}. 

The BLEU score is sensitive to the chosen n-gram order. While a 4-gram model is commonly used in standard evaluations~\cite{denkowski2010choosing, papineni2002bleu}, it may not always be suitable for all contexts. Since the extracted inputs tend to be brief (median is 4), BLEU-2 yields a more accurate reflection of text quality, consistent with approaches used in similar research~\cite{imran2024uncovering}. Therefore, our analysis primarily relies on BLEU-2 for a more precise evaluation of the generated outputs.

\subsection{RQ1: Effectiveness of NLP in Extracting Inputs}
To answer this RQ, we adopt \textit{token classification} approach~\cite{huggingface_token_classification}. In recent years, BERT based transformer models have been widely used for token classification tasks~\cite{chhablani2021nlrg, sun2024lit}. We investigate the effectiveness of BERT in extracting user input commands from unstructured bug reports.

\subsubsection{Dataset}
As our dataset has empty input commands, i.e., bug reports that did not contain any user input commands, we exclude them as they do not contribute to the command extraction task and may introduce noise into the training process. We end up with 149 data instances. The dataset was then split into training (80\%) and testing (20\%) sets.

\subsubsection{Model Configuration}
We use \textit{bert-base-uncased} version. We fine-tune the model for a token classification task by utilizing the Huggingface transformers library~\cite{huggingface}. The primary objective was to perform token classification specifically for extracting user input commands from unstructured bug reports. To achieve this, the model was trained to classify tokens into two categories: \textit{`1'} (insider command) and \textit{`0'} (outside a command). We use AdamW optimizer with a learning rate of 2e-5. The model then trained over ten epochs with a batch size of 16.

\subsubsection{Results}
We observe that the BERT model performed very poorly. The BLEU-2 score in the test set greater than or equal to 0.5 was only once (3.33\%) and in the 0.3-0.5 range only once again (3.33\%), with no exact matches between predictions and ground truth. Manual inspection reveal that 40\% of time the model often  failed to capture a single token in the sequences, and rest 60\% of the time either misclassified command tokens or partially extracted commands. 

We hypothesize that the primary issue is the small size of our dataset, as tasks like command extraction typically require large-scale data~\cite{ma2022extraction, shin2020bert}. Additionally, the model struggled with complex command structures, such as multi-line commands or those embedded within descriptive text, leading to frequent errors.

These results highlight BERT's limitations in handling unstructured bug reports with limited dataset and its inability to generalize to unseen command formats. To address these challenges, we explore \textit{generative language models}~\cite{blease2024generative}, which offer greater flexibility and can operate in zero-shot and one-shot settings, enabling better generalization without extensive fine-tuning.

\subsection{RQ2: Effectiveness of Generative LLMs in Extracting inputs}

To answer this research question, We first design a prompt to extract the input commands from the bug reports. Afterwards, we use the same prompt for all models to extract input commands from bug reports.

\subsubsection{Experiment Setup}

We experiment with two distinct prompting strategies — zero-shot and one-shot — to evaluate the adaptability of LLMs to minimal-context tasks~\cite{zamfirescu2023johnny, dang2022prompt, zhou2024comprehensive}. In the zero-shot setting, models are asked to complete tasks without any examples, relying solely on their pre-trained knowledge. In contrast, the one-shot approach provides a single illustrative example within the prompt to guide the model’s response. After preliminary testing and qualitative assessment, we determined that the one-shot prompt led to more structured and contextually accurate outputs. The finalized prompt design used in the experiments is detailed in section~\ref{prompt}.

We then provide all models with the exact same prompt for each report and collect their responses alongside our manually gathered command sequences.

\subsubsection{Models Configuration} We experiment the performance of three distinct open-source LLMs generative. We particularly focus on open-source models as they are more transparent, customizable, and accessible for research purposes. Below we briefly explain the models:

\textbf{LLaMA:} We utilized the \textit{llama-3.3-70B} model, Meta AI’s latest iteration in the Llama series at the time of this experiment. With its 70 billion parameters, this model excels in both zero-shot and few-shot scenarios, demonstrating advanced reasoning skills and proficiency in handling knowledge-intensive tasks~\cite{touvron2023llama}. Its balanced performance across diverse domains makes it an ideal benchmark for general-purpose tasks.

\textbf{Qwen:}We employed the \textit{qwen2.5:32b-instruct} version, a 32-billion parameter model developed by Alibaba's AI research team. Optimized specifically for instruction-following tasks, Qwen demonstrates a strong ability to comprehend task-oriented prompts and generate structured outputs that align closely with user expectations~\cite{bai2023qwen}. Its design emphasizes prompt adherence and logical task decomposition, making it particularly suitable for procedural tasks~\cite{bai2023qwen}.

\textbf{Qwen-coder:} We incorporated the \textit{qwen2.5-coder:32b} variant, a specialized adaptation of the Qwen model fine-tuned for programming and code-related tasks~\cite{hui2024qwen2}. This variant enhances the model’s capability in generating syntactically accurate and semantically coherent code snippets, making it an ideal candidate for tasks involving software development, code analysis, and algorithmic reasoning~\cite{hui2024qwen2}. 

\subsubsection{Parameters}
As generative LLM's output may vary, we set the temperature 0, so that the output varies least while experimenting. We utilized Ollama platform for accessing these models~\cite{ollama}.

\subsubsection{Prompt Design}

The structure of our input command extraction prompt is intended to
mimic a real-world scenario where a user is going through
bug reports and finding the exact command to replicate the bug report from the command line. The prompt is shown as below:

\begin{promptbox}{Input Command Extractor Prompt}
{
I will provide you with a Linux coreutils bug description that was reported by a developer in Bugzilla. Your task is to determine the exact command(s) or test case(s) required to reproduce the bug. \newline
If a reproducible command or test case exists, write only the command or test case. If no reproducible command or test case is available, write None. \newline

Example: \newline
\textbf{Input (Bug Description)}:
A user reports that the ls command incorrectly sorts filenames when using the -v option. The issue occurs when filenames contain both numbers and letters, leading to unexpected sorting behavior.
\newline
touch file1 file2 file10 file20
ls -v
\newline
\textbf{Output (Expected Response)}:
touch file1 file2 file10 file20
ls -v
\newline

Here is the reported Bug Description:
$<insert\ Bug \ Description>$
\newline
\textbf{Task}: Given the above bug description, identify the commands or test cases required to reproduce the bug.

}
\label{prompt}
\end{promptbox}

\subsubsection{Results}
The range of BLEU scores and their respective count for the three different models using unigram, bigram, trigram, and four-gram are shown in Table~\ref{tab:bleu_scores}. Note here that in some cases both the extracted command and human annotated command were empty, e.g., there were no commands in the bug report. The BLEU score is typically used for non-empty text comparisons. As, theoretically, two empty strings are treated as a perfect match, we consider the BLEU score in such cases 1. 

From Table~\ref{tab:bleu_scores}, we can observe that Qwen performed slightly better than LLaMA and Qwen-Coder. Considering BLEU-2, in 62.62\% cases Qwen has a score $\geq$ 0.5, while LLaMA and Qwen-Coder achieve 55.34\% and 56.8\%. In the case of score $<$ 0.3, Qwen-Coder performs worst, with 38.35\% scoring less than 0.3.

We observe that LLaMA has 48 cases, Qwen has 81 cases and Qwen-Coder has 67 cases where BLEU-2 score is exactly 1, i.e., the human annotation exactly matches with the model extracted outputs. Out of these, in 38 cases all the models have exact matches.

In 81 cases all 3 models have BLEU-score greater than or equal to 0.5. In those 81 cases, the median of the input command to human annotation is 4, and the median of the input command generated by LLaMA, Qwen, and Qwen-Coder are respectively 5, 4, and 4. This is likely because when the models were able to extract correct commands, they were similar to human annotations. On the other hand, in 38 cases all 3 models have BLEU-score less than 0.3. In those 38 cases, the median to human annotation is 4, and the median to LLaMA, Qwen, and Qwen-Coder are respectively 18, 0, and 0. This indicates that when the models were incorrect, a) LLaMA likely extracted a command that were dissimilar to the human annotation, and b) Qwen and Qwen-coder likely predicted `None' in majority of these cases.

\begin{table}[tb]
\centering
\caption{BLEU Score Distribution Across Different Models.}
\label{tab:bleu_scores}
\footnotesize
\renewcommand{\arraystretch}{1.1} 
\begin{tabular} {|p{0.8cm}|p{1cm}|p{1.1cm}|p{1.1cm}|p{1.1cm}|p{1.1cm}|} 
\hline
{Model} & {Range} & {BLEU-1} Count (\%) & \textit{\textbf{BLEU-2}} Count (\%)& {BLEU-3} Count (\%)& {BLEU-4} Count (\%) \\ \hline\hline

 & \multirow{2}{*}{$<$ 0.3} & 74 (35.92\%) & \textit{76 (36.89\%)} & 77 (37.38\%) & 80 (38.83\%) \\ \cline{2-6}
LLaMA & 0.3-$<$0.5 & 14 (6.8\%) & \textit{16 (7.77\%)} & 19 (9.22\%) & 20 (9.71\%) \\ \cline{2-6}
 & \multirow{2}{*}{$\geq$ 0.5} & 118 (57.28\%) & \textit{114 (55.34\%)} & 110 (53.4\%) & 106 (51.46\%) \\ \hline

 & \multirow{2}{*}{$<$ 0.3} &  65 (31.55\%) & \textit{66 (32.04\%)} & 66 (32.04\%) & 69 (33.5\%) \\ \cline{2-6}
Qwen & 0.3-$<$0.5 & 11 (5.34\%) & \textit{11 (5.34\%)} & 15 (7.28\%) & 13 (6.31\%) \\ \cline{2-6}
 & \multirow{2}{*}{$\geq$ 0.5}   & \textbf{130 (63.11\%)} & \textit{\textbf{129 (62.62\%)}} & \textbf{125 (60.68\%)} & \textbf{124 (60.19\%)} \\ \hline

\multirow{5}{=}{Qwen-Coder} 
 & \multirow{2}{*}{$<$ 0.3} &  78 (37.86\%) &\textit{79 (38.35\%)} & 80 (38.83\%) & 82 (39.81\%) \\ \cline{2-6}
 & 0.3-$<$0.5 & 9 (4.37\%) & \textit{10 (4.85\%)} & 11 (5.34\%) & 11 (5.34\%) \\ \cline{2-6}
 & \multirow{2}{*}{$\geq$ 0.5}  & 119 (57.77\%) & \textit{117 (56.8\%)} & 115 (55.83\%) & 113 (54.85\%) \\ 
\hline
\end{tabular}
\end{table}

\subsubsection{Error Analysis}

\begin{table*}[tb]
    \centering
    \caption{Example of Errors}
    \label{tab:errors_in_llama}
    \renewcommand{\arraystretch}{1.2}
    \small
    \begin{tabular}{|p{2cm}|p{5cm}| p{4cm} | p{5cm}|}
        \hline
        \textbf{Error Type} &
        \textbf{Description} & \textbf{Input Command} & \textbf{LLaMA Output} \\
        \hline
        Correct Extraction but Different Wordings
        &
        coreutils-9.3 brought changes to the behavior of the -v option which broke some of my automation scripts [...] Instead they only output this information with --debug [...] &
        1. cp -v -u <source> <destination> \newline 2. mv -v -u <source> <destination> &
        1. cp -u -v file1 file2 \newline 2. mv -i -v file3 file4 \newline 3. cp -u -v --debug file1 file2 \newline 4. mv -i -v --debug file3 file4 \\

        \hline
        Failed Extraction &
        (1) You mounted two partititions \newline 
        (2) You unmount the /mnt (formatted with ext3) \newline
        (3) mount mpath2 (you unmounted in step 2) on /tmp/test/ \newline
        (4) Now , you do df and expect this -  [...] &
        1. mount /dev/mpath1 /tmp/test/umount /mnt \newline 2. mount /dev/mpath2 /tmp/test/df &
        (None) \\

        \hline
        Subjectivity in Annotation & command install sets wrong permissions [..] \newline
        Actual results: -rwxr-xr-x 1 root [..] \newline
        Expected results: -rw-r--r-- 1 root [..] &
        (None) &
        1. bash install 
        
        /home/vikram/Desktop/RPM-GPG-KEY-AL-CAPCOM /etc/pki/rpm-gpg/[...] \\
        \hline
    \end{tabular}
\end{table*}

To understand the models' errors, we analyzed 38 instances where all three models had a BLEU score below 0.30. We focus particularly on the LLaMA model, as it produced fewer `None' predictions, resulting in more samples with interpretable text. 

Two authors carefully examined the human-annotated input commands and the LLM-generated commands. They employed thematic analysis to categorize the errors~\cite{saunders2018saturation}. Specifically, One author first examined the commands and identified initial themes, which were then evaluated and refined by another author. They collaboratively discussed any inconsistencies and reached a consensus to finalize the categorization. Our analysis identifies three main error categories:

\smallskip
\noindent
\textbf{Correct Extraction but Different Wordings}: There were 25 cases where the models generated slightly different variations of the input command, rather than the exact wordings. In those cases, while the models are technically correct, their generated command wordings were different. 

\smallskip
\noindent
\textbf{Failed Extraction}: There are 10 cases where the models failed to extract the input commands or test cases from the bug report where the annotators perceived otherwise.

\smallskip
\noindent
\textbf{Subjectivity in Annotation}: There were 3 cases where it was not clear if the input commands or test cases were actually in the bug report. Those cases where the model suggested that there were likely input commands, however, annotators were uncertain. Those are the cases, which further needed to be verified by the domain experts.

Table~\ref{tab:errors_in_llama} shows the examples of these errors.

\section{Discussion and Future Scope}




Our error analysis reveals several avenues for enhancing the extraction of input commands from bug reports. Although our study relies primarily on BLEU scores, future work should integrate additional metrics—such as semantic similarity and context-aware measures—to capture subtle phrasing variations and provide a more comprehensive evaluation. Refining prompt design with dynamic, multi-stage approaches that incorporate contextual feedback may further align model outputs with human annotations.

Another promising direction is the integration of domain-specific knowledge. Our experiments indicate that models sometimes struggle with implicit or context-dependent commands; thus, fine-tuning on specialized datasets from diverse software domains could enhance their understanding of command patterns and improve extraction accuracy.

Finally, addressing the challenges of annotation subjectivity, where the presence of commands is ambiguous, may benefit from continuous expert feedback during training and evaluation. Broadening the evaluation framework to encompass bug reports from different domains and languages—and exploring hybrid approaches that blend rule-based heuristics with LLM outputs—will likely lead to more robust, context-aware command extraction systems and more effective debugging in real-world software development environments.

\section{Related Work}

Our work builds upon and extends previous research in two main areas: extracting structured information and how Prompt-based techniques for information extraction from bug reports. Below we discuss them.

\noindent\textbf{Extracting Structured Information from Bug Reports:} Extracting information from bug report has been a matured research area~\cite{bettenburg2008extracting, zhang2015survey}. Prior work has leveraged neural language techniques for bug report analysis, developing automated tools and frameworks that enhance bug identification and reduce manual effort significantly.
Bettenburg identified which elements in bug reports are most valuable to developers~\cite{bettenburg2008extracting}.
Key advances include Rastkar et al.'s and Mani et al.'s work on automatic summarization of bug reports~\cite{rastkar2014automatic, mani2012ausum}.
Chaparro et al. developed methods for detecting missing information in bug descriptions~\cite{chaparro2017detecting, chaparro2019assessing}, while Imran et al. introduced techniques for finding missing information from bug reports~\cite{imran2021automatically}. Feng et al. analyzed secure device vulnerabilities through NLP-based techniques~ \cite{feng2019understanding}. Zhu et al. proposed a graph-based neural model for automated bug localization with bug report decomposition and code hierarchical network~\cite{zhu2022enhancing}. Zhao et al. developed a tool that extract S2R from bug reports~\cite{zhao2019automatically}. Ciborowska et al. proposed a BERT-based technique for bug localization~\cite{ciborowska2022fast}. Ouedraogo et al. proposed an appraoch for automatic test case generation after extracting relevant information from bug reports~\cite{ouedraogo2023enriching}.

\noindent
\textbf{Information Extraction from Bug Reports Through Prompting:}
Recent advances in generative LLMs have introduced prompt-based techniques for bug report analysis. 
Feng et al. proposed a technique to automatically reproduce the bugs from bug reports through prompt engineering~\cite{feng2024prompting}. Plein did a feasibility study with LLMs on test case generation from bug reports~\cite{plein2024automatic}. Wen et al. analyzed static bug report warnings~\cite{wen2024automatically}. Ahmed et al. experimented on how capable GPT-4 and GPT-3.5 are to fix hardware based security bugs~\cite{ahmad2024hardware}. Kang et al. proposed a prompt based framework for bug reproduction~\cite{kang2023large}. Guan et al. introduced a knowledge-aware prompting framework that utilizes LLMs to generate test cases for model optimization bug~\cite{guan2024large}. Paul et al. applied zero-shot prompting technique using GPT-3.5-turbo for automatic program repair and found that they don't perform well~\cite{paul2023enhancing}.

Unlike those works, our empirical study specifically focus on how good \textit{Open-source LLMs} are on extracting \textit{input commands} from Linux Coreutils bug reports.

\section{Threats to Validity}
In this section, we discuss four types of threats, similar to prior research related to LLM\cite{pyreddy2025emoxptanalyzingemotionalvariances, hossain2025llmprosanalyzinglargelanguage}.

\noindent
\textbf{Threats to Internal Validity:}
Data contamination significantly threatens the internal validity of our study, as some of the bug reports we utilized may already be present in the training data of the evaluated LLMs.

\noindent
\textbf{Threats to External Validity:}
The study focuses on bug reports from Linux coreutils, specifically those reported on Redhat Bugzilla, which may limit the generalizability of our findings. While the Linux coreutils bug reports are diverse and complex, they represent only a subset of web application bug reports. Additionally, different websites may have distinct structures for bug reporting. In this study, we analyze bug reports from just one platform (Redhat), which could affect the performance of LLMs and limit the applicability of our conclusions to other domains of bug reports. To address this limitation, we include a diverse set of 250 bug reports, spanning various categories and difficulty levels, to strengthen the robustness of our evaluation.

\noindent
\textbf{Threats to Construct Validity:}
We test the models in a one-shot setting, without additional fine-tuning or iterative interactions. This approach may not fully utilize the models’ capabilities, as fine-tuning or interactive prompting could enhance performance on specific tasks. To address this limitation, we document all experimental procedures and make the scripts used for data preprocessing and solution evaluation publicly available, ensuring reproducibility and transparency.

\noindent
\textbf{Threats to Conclusion Validity:}
Our evaluation relies on the data annotation phase and the prompt used to assess the performance of the LLM. To minimize bias, we ensure that all evaluations are conducted using the same prompt. Additionally, we verify the manual data annotation by having two different authors independently review the annotations.

\section{Conclusion}
Our investigation of \Name{} demonstrates that generative LLMs can effectively extract failure-inducing inputs from bug reports without extensive training data compared to BERT-based model, which performed poorly. The generative models showed promising results, with Qwen achieving the highest accuracy (62.62\% outputs with BLEU-2 $\geq$ 0.5). Key challenges identified include command wording variations, extraction failures, and annotation ambiguity. These findings suggest significant potential for integrating LLMs into bug reproduction workflows, reducing manual effort for developers. Future work should focus on domain-specific knowledge integration, refined prompting strategies, and hybrid approaches to further improve extraction accuracy and reliability across diverse software ecosystems.
\section{Acknowledgments}
This work was supported in part by NSF grants CCF-2348277 and CCF-2518445.

\bibliographystyle{ACM-Reference-Format}  
\bibliography{references}

@article{touvron2023llama,
  title={Llama: Open and efficient foundation language models},
  author={Touvron, Hugo and Lavril, Thibaut and Izacard, Gautier and Martinet, Xavier and Lachaux, Marie-Anne and Lacroix, Timoth{\'e}e and Rozi{\`e}re, Baptiste and Goyal, Naman and Hambro, Eric and Azhar, Faisal and others},
  journal={arXiv preprint arXiv:2302.13971},
  year={2023}
}

@article{bai2023qwen,
  title={Qwen technical report},
  author={Bai, Jinze and Bai, Shuai and Chu, Yunfei and Cui, Zeyu and Dang, Kai and Deng, Xiaodong and Fan, Yang and Ge, Wenbin and Han, Yu and Huang, Fei and others},
  journal={arXiv preprint arXiv:2309.16609},
  year={2023}
}

@article{hui2024qwen2,
  title={Qwen2. 5-coder technical report},
  author={Hui, Binyuan and Yang, Jian and Cui, Zeyu and Yang, Jiaxi and Liu, Dayiheng and Zhang, Lei and Liu, Tianyu and Zhang, Jiajun and Yu, Bowen and Lu, Keming and others},
  journal={arXiv preprint arXiv:2409.12186},
  year={2024}
}

@inproceedings{feng2024prompting,
  title={Prompting is all you need: Automated android bug replay with large language models},
  author={Feng, Sidong and Chen, Chunyang},
  booktitle={Proceedings of the 46th IEEE/ACM International Conference on Software Engineering},
  pages={1--13},
  year={2024}
}

@article{zhu2022enhancing,
  title={Enhancing bug localization with bug report decomposition and code hierarchical network},
  author={Zhu, Ziye and Tong, Hanghang and Wang, Yu and Li, Yun},
  journal={Knowledge-Based Systems},
  volume={248},
  pages={108741},
  year={2022},
  publisher={Elsevier}
}

@article{ahmad2024hardware,
  title={On hardware security bug code fixes by prompting large language models},
  author={Ahmad, Baleegh and Thakur, Shailja and Tan, Benjamin and Karri, Ramesh and Pearce, Hammond},
  journal={IEEE Transactions on Information Forensics and Security},
  year={2024},
  publisher={IEEE}
}

@article{paul2023enhancing,
  title={Enhancing automated program repair through fine-tuning and prompt engineering},
  author={Paul, Rishov and Hossain, Md Mohib and Siddiq, Mohammed Latif and Hasan, Masum and Iqbal, Anindya and Santos, Joanna},
  journal={arXiv preprint arXiv:2304.07840},
  year={2023}
}

@inproceedings{guan2024large,
  title={Large language models can connect the dots: Exploring model optimization bugs with domain knowledge-aware prompts},
  author={Guan, Hao and Bai, Guangdong and Liu, Yepang},
  booktitle={Proceedings of the 33rd ACM SIGSOFT International Symposium on Software Testing and Analysis},
  pages={1579--1591},
  year={2024}
}

@article{wen2024automatically,
  title={Automatically inspecting thousands of static bug warnings with large language model: How far are we?},
  author={Wen, Cheng and Cai, Yuandao and Zhang, Bin and Su, Jie and Xu, Zhiwu and Liu, Dugang and Qin, Shengchao and Ming, Zhong and Cong, Tian},
  journal={ACM Transactions on Knowledge Discovery from Data},
  volume={18},
  number={7},
  pages={1--34},
  year={2024},
  publisher={ACM New York, NY}
}

@inproceedings{plein2024automatic,
  title={Automatic generation of test cases based on bug reports: a feasibility study with large language models},
  author={Plein, Laura and Ou{\'e}draogo, Wendk{\^u}uni C and Klein, Jacques and Bissyand{\'e}, Tegawend{\'e} F},
  booktitle={Proceedings of the 2024 IEEE/ACM 46th International Conference on Software Engineering: Companion Proceedings},
  pages={360--361},
  year={2024}
}

@inproceedings{ouedraogo2023enriching,
  title={Extracting Relevant Test Inputs from Bug Reports for Automatic Test Case Generation},
  author={Ou{\'e}draogo, Wendk{\^u}uni C and Plein, Laura and Kabor{\'e}, Kader and Habib, Andrew and Klein, Jacques and Lo, David and Bissyand{\'e}, Tegawend{\'e} F},
  booktitle={Proceedings of the 2024 IEEE/ACM 46th International Conference on Software Engineering: Companion Proceedings},
  pages={406--407},
  year={2024}
}

@BOOK{Adreas:Zeller,
  TITLE = "Why Programs Fail: A Guide to Systematic Debugging",
  SUBTITLE = "",
  AUTHOR = "Andreas Zeller",
  year = "October 2005", 
  PUBLISHER = ""
}

@inproceedings{10.1145/2652524.2652541,
author = {Davies, Steven and Roper, Marc},
title = {What's in a bug report?},
year = {2014},
isbn = {9781450327749},
publisher = {Association for Computing Machinery},
address = {New York, NY, USA},
url = {https://doi.org/10.1145/2652524.2652541},
doi = {10.1145/2652524.2652541},
booktitle = {Proceedings of the 8th ACM/IEEE International Symposium on Empirical Software Engineering and Measurement},
articleno = {26},
numpages = {10},
location = {Torino, Italy},
series = {ESEM '14}
}

@INPROCEEDINGS{8811942,
  author={Zhao, Yu and Yu, Tingting and Su, Ting and Liu, Yang and Zheng, Wei and Zhang, Jingzhi and G.J. Halfond, William},
  booktitle={2019 IEEE/ACM 41st International Conference on Software Engineering (ICSE)}, 
  title={ReCDroid: Automatically Reproducing Android Application Crashes from Bug Reports}, 
  year={2019},
  volume={},
  number={},
  pages={128-139},
  doi={10.1109/ICSE.2019.00030}}

@inproceedings{imran2021automatically,
  title={Automatically selecting follow-up questions for deficient bug reports},
  author={Imran, Mia Mohammad and Ciborowska, Agnieszka and Damevski, Kostadin},
  booktitle={2021 IEEE/ACM 18th International Conference on Mining Software Repositories (MSR)},
  pages={167--178},
  year={2021},
  organization={IEEE}
}

@inproceedings{chaparro2019assessing,
  title={Assessing the quality of the steps to reproduce in bug reports},
  author={Chaparro, Oscar and Bernal-C{\'a}rdenas, Carlos and Lu, Jing and Moran, Kevin and Marcus, Andrian and Di Penta, Massimiliano and Poshyvanyk, Denys and Ng, Vincent},
  booktitle={Proceedings of the 2019 27th ACM joint meeting on european software engineering conference and symposium on the foundations of software engineering},
  pages={86--96},
  year={2019}
}

@inproceedings{chaparro2017detecting,
  title={Detecting missing information in bug descriptions},
  author={Chaparro, Oscar and Lu, Jing and Zampetti, Fiorella and Moreno, Laura and Di Penta, Massimiliano and Marcus, Andrian and Bavota, Gabriele and Ng, Vincent},
  booktitle={Proceedings of the 2017 11th joint meeting on foundations of software engineering},
  pages={396--407},
  year={2017}
}

@inproceedings{ciborowska2022fast,
  title={Fast changeset-based bug localization with bert},
  author={Ciborowska, Agnieszka and Damevski, Kostadin},
  booktitle={Proceedings of the 44th international conference on software engineering},
  pages={946--957},
  year={2022}
}

@article{rastkar2014automatic,
  title={Automatic summarization of bug reports},
  author={Rastkar, Sarah and Murphy, Gail C and Murray, Gabriel},
  journal={IEEE Transactions on Software Engineering},
  volume={40},
  number={4},
  pages={366--380},
  year={2014},
  publisher={IEEE}
}

@inproceedings{10.1145/3106237.3106266,
author = {Yu, Tingting and Zaman, Tarannum S. and Wang, Chao},
title = {DESCRY: reproducing system-level concurrency failures},
year = {2017},
isbn = {9781450351058},
publisher = {Association for Computing Machinery},
address = {New York, NY, USA},
url = {https://doi.org/10.1145/3106237.3106266},
doi = {10.1145/3106237.3106266},
booktitle = {Proceedings of the 2017 11th Joint Meeting on Foundations of Software Engineering},
pages = {694–704},
numpages = {11},
location = {Paderborn, Germany},
series = {ESEC/FSE 2017}
}

@inproceedings{feng2019understanding,
  title={Understanding and securing device vulnerabilities through automated bug report analysis},
  author={Feng, Xuan and Liao, Xiaojing and Wang, X and Wang, Haining and Li, Qiang and Yang, Kai and Zhu, Hongsong and Sun, Limin},
  booktitle={SEC'19: Proceedings of the 28th USENIX Conference on Security Symposium},
  year={2019}
}

@inproceedings{mani2012ausum,
  title={Ausum: approach for unsupervised bug report summarization},
  author={Mani, Senthil and Catherine, Rose and Sinha, Vibha Singhal and Dubey, Avinava},
  booktitle={Proceedings of the ACM SIGSOFT 20th International Symposium on the Foundations of Software Engineering},
  pages={1--11},
  year={2012}
}

@inproceedings{zhao2019automatically,
  title={Automatically extracting bug reproducing steps from android bug reports},
  author={Zhao, Yu and Miller, Kye and Yu, Tingting and Zheng, Wei and Pu, Minchao},
  booktitle={Reuse in the Big Data Era: 18th International Conference on Software and Systems Reuse, ICSR 2019, Cincinnati, OH, USA, June 26--28, 2019, Proceedings 18},
  pages={100--111},
  year={2019},
  organization={Springer}
}

@article{zhang2015survey,
  title={A survey on bug-report analysis.},
  author={Zhang, Jie and Wang, Xiaoyin and Hao, Dan and Xie, Bing and Zhang, Lu and Mei, Hong},
  journal={Sci. China Inf. Sci.},
  volume={58},
  number={2},
  pages={1--24},
  year={2015},
  publisher={Citeseer}
}

@inproceedings{bettenburg2008extracting,
  title={Extracting structural information from bug reports},
  author={Bettenburg, Nicolas and Premraj, Rahul and Zimmermann, Thomas and Kim, Sunghun},
  booktitle={Proceedings of the 2008 international working conference on Mining software repositories},
  pages={27--30},
  year={2008}
}

@inproceedings{kang2023large,
  title={Large language models are few-shot testers: Exploring llm-based general bug reproduction},
  author={Kang, Sungmin and Yoon, Juyeon and Yoo, Shin},
  booktitle={2023 IEEE/ACM 45th International Conference on Software Engineering (ICSE)},
  pages={2312--2323},
  year={2023},
  organization={IEEE}
}

@misc{hossain2025llmprosanalyzinglargelanguage,
      title={LLM-ProS: Analyzing Large Language Models' Performance in Competitive Problem Solving}, 
      author={Md Sifat Hossain and Anika Tabassum and Md. Fahim Arefin and Tarannum Shaila Zaman},
      year={2025},
      eprint={2502.04355},
      archivePrefix={arXiv},
      primaryClass={cs.CL},
      url={https://arxiv.org/abs/2502.04355}, 
}

@misc{pyreddy2025emoxptanalyzingemotionalvariances,
      title={EmoXpt: Analyzing Emotional Variances in Human Comments and LLM-Generated Responses}, 
      author={Shireesh Reddy Pyreddy and Tarannum Shaila Zaman},
      year={2025},
      eprint={2501.06597},
      archivePrefix={arXiv},
      primaryClass={cs.LG},
      url={https://arxiv.org/abs/2501.06597}, 
}

@online{redhat,
  key =          {redhat$\_$bug},
  title =        "Red Hat Bugzilla – Main Page",
  url =          "https://bugzilla.redhat.com/",
  lastaccessed = "June 01, 2020",
}

@online{coreutils,
  key =          {coreutils},
  title =        "Decoded: GNU coreutils",
  url =          "https://www.maizure.org/projects/decoded-gnu-coreutils/",
  lastaccessed = "Feb. 24, 2025",
}

@article{saunders2018saturation,
  title={Saturation in qualitative research: exploring its conceptualization and operationalization},
  author={B. Saunders and J. Sim and T. Kingstone and S. Baker and J. Waterfield and B. Bartlam and H. Burroughs and C. Jinks},
  journal={Quality \& quantity},
  volume={52},
  year={2018},
  publisher={Springer}
}

@phdthesis{Casey:1992:CPD:144167,
 author = {Casey, Patrick Joseph},
 title = {Computer Program Debugging: An Engaging Problem-solving Environment},
 year = {1992},
 note = {UMI Order No. GAX92-10525},
 publisher = {Columbia University Teacher's College},
 address = {New York, NY, USA},
}

@inproceedings{papineni2002bleu,
  title={Bleu: a method for automatic evaluation of machine translation},
  author={Papineni, Kishore and Roukos, Salim and Ward, Todd and Zhu, Wei-Jing},
  booktitle={Proceedings of the 40th annual meeting of the Association for Computational Linguistics},
  pages={311--318},
  year={2002}
}

@inproceedings{denkowski2010choosing,
  title={Choosing the right evaluation for machine translation: an examination of annotator and automatic metric performance on human judgment tasks},
  author={Denkowski, Michael and Lavie, Alon},
  booktitle={Proceedings of the 9th Conference of the Association for Machine Translation in the Americas: Research Papers},
  year={2010}
}

@inproceedings{zamfirescu2023johnny,
  title={Why Johnny can’t prompt: how non-AI experts try (and fail) to design LLM prompts},
  author={Zamfirescu-Pereira, J Diego and Wong, Richmond Y and Hartmann, Bjoern and Yang, Qian},
  booktitle={Proceedings of the 2023 CHI conference on human factors in computing systems},
  pages={1--21},
  year={2023}
}

@inproceedings{deng2024information,
  title={Information extraction in low-resource scenarios: Survey and perspective},
  author={Deng, Shumin and Ma, Yubo and Zhang, Ningyu and Cao, Yixin and Hooi, Bryan},
  booktitle={2024 IEEE International Conference on Knowledge Graph (ICKG)},
  pages={33--49},
  year={2024},
  organization={IEEE}
}

@article{xu2024large,
  title={Large language models for generative information extraction: A survey},
  author={Xu, Derong and Chen, Wei and Peng, Wenjun and Zhang, Chao and Xu, Tong and Zhao, Xiangyu and Wu, Xian and Zheng, Yefeng and Wang, Yang and Chen, Enhong},
  journal={Frontiers of Computer Science},
  volume={18},
  number={6},
  pages={186357},
  year={2024},
  publisher={Springer}
}

@incollection{alaofi2024generative,
  title={Generative information retrieval evaluation},
  author={Alaofi, Marwah and Arabzadeh, Negar and Clarke, Charles LA and Sanderson, Mark},
  booktitle={Information Access in the Era of Generative AI},
  pages={135--159},
  year={2024},
  publisher={Springer}
}

@inproceedings{li2023codeie,
  title={CodeIE: Large Code Generation Models are Better Few-Shot Information Extractors},
  author={Li, Peng and Sun, Tianxiang and Tang, Qiong and Yan, Hang and Wu, Yuanbin and Huang, Xuan-Jing and Qiu, Xipeng},
  booktitle={Proceedings of the 61st Annual Meeting of the Association for Computational Linguistics (Volume 1: Long Papers)},
  pages={15339--15353},
  year={2023}
}

@inproceedings{sainz2023gollie,
  title={GoLLIE: Annotation Guidelines improve Zero-Shot Information-Extraction},
  author={Sainz, Oscar and Garc{\'\i}a-Ferrero, Iker and Agerri, Rodrigo and de Lacalle, Oier Lopez and Rigau, German and Agirre, Eneko},
  booktitle={ICLR},
  year={2024}
}

@inproceedings{mo2024c,
  title={C-ICL: Contrastive In-context Learning for Information Extraction},
  author={Mo, Ying and Liu, Jiahao and Yang, Jian and Wang, Qifan and Zhang, Shun and Wang, Jingang and Li, Zhoujun},
  booktitle={Findings of the Association for Computational Linguistics: EMNLP 2024},
  pages={10099--10114},
  year={2024}
}

@inproceedings{zhang2025survey,
  title={A Survey of Generative Information Extraction},
  author={Zhang, Zikang and You, Wangjie and Wu, Tianci and Wang, Xinrui and Li, Juntao and Zhang, Min},
  booktitle={Proceedings of the 31st International Conference on Computational Linguistics},
  pages={4840--4870},
  year={2025}
}

@article{blease2024generative,
  title={Generative language models and open notes: exploring the promise and limitations},
  author={Blease, Charlotte and Torous, John and McMillan, Brian and H{\"a}gglund, Maria and Mandl, Kenneth D},
  journal={JMIR medical education},
  volume={10},
  pages={e51183},
  year={2024},
  publisher={JMIR Publications Toronto, Canada}
}

@article{zhou2024comprehensive,
  title={A comprehensive survey on pretrained foundation models: A history from bert to chatgpt},
  author={Zhou, Ce and Li, Qian and Li, Chen and Yu, Jun and Liu, Yixin and Wang, Guangjing and Zhang, Kai and Ji, Cheng and Yan, Qiben and He, Lifang and others},
  journal={International Journal of Machine Learning and Cybernetics},
  pages={1--65},
  year={2024},
  publisher={Springer}
}

@article{dang2022prompt,
  title={How to prompt? Opportunities and challenges of zero-and few-shot learning for human-AI interaction in creative applications of generative models},
  author={Dang, Hai and Mecke, Lukas and Lehmann, Florian and Goller, Sven and Buschek, Daniel},
  journal={arXiv preprint arXiv:2209.01390},
  year={2022}
}

@inproceedings{lavie-2011-evaluating,
    title = "Evaluating the Output of Machine Translation Systems",
    author = "Lavie, Alon",
    booktitle = "Proceedings of Machine Translation Summit XIII: Tutorial Abstracts",
    month = sep # " 19",
    year = "2011",
    address = "Xiamen, China",
    url = "https://aclanthology.org/2011.mtsummit-tutorials.3/"
}

@misc{google_translate_automl_evaluate,
  author       = {Google Cloud},
  title        = {AutoML Evaluation | Cloud Translation | Google Cloud},
  year         = {2025},
  url          = {https://cloud.google.com/translate/docs/advanced/automl-evaluate}
}

@inproceedings{Seljan2012BLEUEO,
    title = "{BLEU} Evaluation of Machine-Translated {E}nglish-{C}roatian Legislation",
    author = "Seljan, Sanja  and
      Brki{\'c}, Marija  and
      Vi{\v{c}}i{\'c}, Tomislav",
    booktitle = "Proceedings of the Eighth International Conference on Language Resources and Evaluation ({LREC}'12)",
    month = may,
    year = "2012",
    address = "Istanbul, Turkey",
    pages = "2143--2148",
}

@article{gao2020generating,
  title={Generating question titles for stack overflow from mined code snippets},
  author={Gao, Zhipeng and Xia, Xin and Grundy, John and Lo, David and Li, Yuan-Fang},
  journal={ACM Transactions on Software Engineering and Methodology (TOSEM)},
  volume={29},
  number={4},
  pages={1--37},
  year={2020},
  publisher={ACM New York, NY, USA}
}

@inproceedings{imran2024uncovering,
  title={Uncovering the causes of emotions in software developer communication using zero-shot llms},
  author={Imran, Mia Mohammad and Chatterjee, Preetha and Damevski, Kostadin},
  booktitle={Proceedings of the IEEE/ACM 46th international conference on software engineering},
  pages={1--13},
  year={2024}
}

@inproceedings{gu2023llm,
  title={Llm-based code generation method for golang compiler testing},
  author={Gu, Qiuhan},
  booktitle={Proceedings of the 31st ACM Joint European Software Engineering Conference and Symposium on the Foundations of Software Engineering},
  pages={2201--2203},
  year={2023}
}

@inproceedings{akuthota2023vulnerability,
  title={Vulnerability detection and monitoring using llm},
  author={Akuthota, Vishwanath and Kasula, Raghunandan and Sumona, Sabiha Tasnim and Mohiuddin, Masud and Reza, Md Tanzim and Rahman, Md Mizanur},
  booktitle={2023 IEEE 9th International Women in Engineering (WIE) Conference on Electrical and Computer Engineering (WIECON-ECE)},
  pages={309--314},
  year={2023},
  organization={IEEE}
}

@article{lu2024grace,
  title={GRACE: Empowering LLM-based software vulnerability detection with graph structure and in-context learning},
  author={Lu, Guilong and Ju, Xiaolin and Chen, Xiang and Pei, Wenlong and Cai, Zhilong},
  journal={Journal of Systems and Software},
  volume={212},
  pages={112031},
  year={2024},
  publisher={Elsevier}
}

@article{hossain2024deep,
  title={A deep dive into large language models for automated bug localization and repair},
  author={Hossain, Soneya Binta and Jiang, Nan and Zhou, Qiang and Li, Xiaopeng and Chiang, Wen-Hao and Lyu, Yingjun and Nguyen, Hoan and Tripp, Omer},
  journal={Proceedings of the ACM on Software Engineering},
  volume={1},
  number={FSE},
  pages={1471--1493},
  year={2024},
  publisher={ACM New York, NY, USA}
}

@inproceedings{yang2024large,
  title={Large language models for test-free fault localization},
  author={Yang, Aidan ZH and Le Goues, Claire and Martins, Ruben and Hellendoorn, Vincent},
  booktitle={Proceedings of the 46th IEEE/ACM International Conference on Software Engineering},
  pages={1--12},
  year={2024}
}

@inproceedings{ahmed2024automatic,
  title={Automatic semantic augmentation of language model prompts (for code summarization)},
  author={Ahmed, Toufique and Pai, Kunal Suresh and Devanbu, Premkumar and Barr, Earl},
  booktitle={Proceedings of the IEEE/ACM 46th international conference on software engineering},
  pages={1--13},
  year={2024}
}

@inproceedings{ahmed2022few,
  title={Few-shot training llms for project-specific code-summarization},
  author={Ahmed, Toufique and Devanbu, Premkumar},
  booktitle={Proceedings of the 37th IEEE/ACM international conference on automated software engineering},
  pages={1--5},
  year={2022}
}

@article{ma2022extraction,
  title={Extraction of temporal information from social media messages using the BERT model},
  author={Ma, Kai and Tan, Yongjian and Tian, Miao and Xie, Xuejing and Qiu, Qinjun and Li, Sanfeng and Wang, Xin},
  journal={Earth Science Informatics},
  volume={15},
  number={1},
  pages={573--584},
  year={2022},
  publisher={Springer}
}

@inproceedings{shin2020bert,
  title={BERT-based spatial information extraction},
  author={Shin, Hyeong Jin and Park, Jeong Yeon and Yuk, Dae Bum and Lee, Jae Sung},
  booktitle={Proceedings of the Third International Workshop on Spatial Language Understanding},
  pages={10--17},
  year={2020}
}

@inproceedings{sun2024lit,
  title={LIT: Label-Informed Transformers on Token-Based Classification},
  author={Sun, Wenjun and Tran, Hanh Thi Hong and Gonz{\'a}lez-Gallardo, Carlos-Emiliano and Coustaty, Micka{\"e}l and Doucet, Antoine},
  booktitle={International Conference on Theory and Practice of Digital Libraries},
  pages={144--158},
  year={2024},
  organization={Springer}
}

@inproceedings{chhablani2021nlrg,
  title={NLRG at SemEval-2021 Task 5: Toxic Spans Detection Leveraging BERT-based Token Classification and Span Prediction Techniques},
  author={Chhablani, Gunjan and Sharma, Abheesht and Pandey, Harshit and Bhartia, Yash and Suthaharan, Shan},
  booktitle={Proceedings of the 15th International Workshop on Semantic Evaluation (SemEval-2021)},
  pages={233--242},
  year={2021}
}

@inproceedings{10.1145/3476883.3520207,
author = {Zaman, Tarannum Shaila and Islam, Tariqul},
title = {ReDPro: an automated technique to detect and regenerate process-level concurrency failures},
year = {2022},
isbn = {9781450386975},
publisher = {Association for Computing Machinery},
address = {New York, NY, USA},
url = {https://doi.org/10.1145/3476883.3520207},
doi = {10.1145/3476883.3520207},
booktitle = {Proceedings of the 2022 ACM Southeast Conference},
pages = {106–112},
numpages = {7},
location = {Virtual Event},
series = {ACMSE '22}
}

@article{10.1145/347636.348938,
author = {Hildebrandt, Ralf and Zeller, Andreas},
title = {Simplifying failure-inducing input},
year = {2000},
issue_date = {Sept. 2000},
publisher = {Association for Computing Machinery},
address = {New York, NY, USA},
volume = {25},
number = {5},
issn = {0163-5948},
url = {https://doi.org/10.1145/347636.348938},
doi = {10.1145/347636.348938},
journal = {SIGSOFT Softw. Eng. Notes},
month = aug,
pages = {135–145},
numpages = {11}
}

@inproceedings{du2024llmsassistnlpresearchers,
  title={LLMs Assist NLP Researchers: Critique Paper (Meta-) Reviewing},
  author={Du, Jiangshu and Wang, Yibo and Zhao, Wenting and Deng, Zhongfen and Liu, Shuaiqi and Lou, Renze and Zou, Henry and Venkit, Pranav Narayanan and Zhang, Nan and Srinath, Mukund and others},
  booktitle={Proceedings of the 2024 Conference on Empirical Methods in Natural Language Processing},
  pages={5081--5099},
  year={2024}
}

@misc{zaman2025genderdynamicssoftwareengineering,
      title={Gender Dynamics in Software Engineering: Insights from Research on Concurrency Bug Reproduction}, 
      author={Tarannum Shaila Zaman and Macharla Hemanth Kishan and Lutfun Nahar Lota},
      year={2025},
      eprint={2502.20289},
      archivePrefix={arXiv},
      primaryClass={cs.SE},
      url={https://arxiv.org/abs/2502.20289}, 
}

@article{lota5060080recent,
  title={Recent Trends and Challenges in Using Nlp Techniques in Software Debugging: A Systematic Literature Review},
  author={Lota, Lutfun Nahar and Zaman, Tarannum Shaila and Azwad, Mirza Mohammad and Farah, Labiba and Chowdhury, Abrar and Anjum, Zaarin and Islam, Chadni and Kamal, Abu Raihan Mostofa},
  journal={Available at SSRN 5060080}
}

@misc{huggingface_token_classification,
  author       = {Hugging Face},
  title        = {Token Classification - Hugging Face},
  year         = {2025},
  url          = {https://huggingface.co/tasks/token-classification}
}

@misc{ollama,
  author       = {ollama.com},
  title        = {Ollama},
  year         = {2025},
  url          = {https://ollama.com/}
}

@misc{huggingface,
  author       = {Hugging Face},
  title        = {Hugging Face},
  year         = {2025},
  url          = {https://huggingface.co}
}

@proceedings{dataset,
  title        = {LLPut: Investigating Large Language Models for Bug
                   Report-Based Input Generation
                  },
  year         = 2025,
  publisher    = {Zenodo},
  month        = mar,
  doi          = {10.5281/zenodo.15092886},
  url          = {https://doi.org/10.5281/zenodo.15092886},
}

\end{sloppypar}
\end{document}